\documentclass[12pt]{iopart}
\usepackage{graphicx,color}
\begin{document}
\title[Experimental evidence of ageing and slow restoration of the weak-contact configuration\ldots]{Experimental evidence of ageing and slow restoration of the weak-contact configuration in tilted 3D granular packings}
\author{S. Kiesgen de Richter$^{1,\dagger}$, V. Yu. Zaitsev$^{2}$, P. Richard$^{1}$, R. Delannay$^{1}$, G. Le Ca{\"e}r$^{1}$ and V. Tournat$^3$}

%\author{S. Kiesgen de Richter}%
% \email{Second.Author@institution.edu.}

\address{$^1$  
Institut de Physique de Rennes, UMR CNRS 6251, Universit\'e de Rennes 1 - 263 av. G\'en\'eral Leclerc, 
35042, Rennes, France\\%\\This line break forced with \textbackslash\textbackslash
}
\address{$^2$  
Institute of Applied Physics, Russian Academy of Sciences, Uljanova St. 46, 603950, Nizhny Novgorod, Russia%\\This line break forced with \textbackslash\textbackslash
}%
\address{$^3$
LAUM, CNRS, Universit\'e du Maine - av. Olivier Messiaen, 72085, Le Mans Cedex 9, France%\\This line break forced% with \\
}%
\address{$^\dagger$ Present address: Laboratoire d'\'Energ\'etique et de M\'ecanique Th\'eorique et Appliqu\'ee, UMR CNRS 7563, 2 avenue de la for\^ et de Haye, BP 160, F-54504 Vand\oe uvre-les-Nancy, France}
\date{\today}% It is always \today, today,
             %  but any date may be explicitly specified
\ead{sebastien.kiesgen@ensem.inpl-nancy.fr, vyuzai@hydro.appl.sci-nnov.ru, patrick.richard@univ-rennes1.fr, renaud.delannay@univ-rennes1.fr, vincent.tournat@univ-lemans.fr, gerard.le-caer@univ-rennes1.fr}
\begin{abstract}
Granular packings slowly driven towards their instability threshold are studied using a digital
imaging technique as well as a nonlinear acoustic method. The former method allows us to study grain rearrangements on the surface during the tilting and the latter enables to selectively probe the modifications of the weak-contact fraction in the material bulk. 
Gradual ageing of both the surface activity and the weak-contact reconfigurations is observed as a result of repeated tilt cycles up to a given angle smaller than the angle of avalanche. 
For an aged configuration reached after several consecutive tilt cycles, abrupt resumption of the on-surface activity and of the weak-contact rearrangements occurs when the packing is subsequently inclined beyond the previous maximal tilting angle. This behavior is compared with literature results from numerical simulations of inclined 2D packings. It is also found that the aged weak-contact configurations exhibit spontaneous restoration towards the initial state if the packing remains at rest for tens of minutes. When the packing is titled forth and back between zero and near-critical angles, instead of ageing, the weak-contact configuration exhibits "internal weak-contact avalanches" in the vicinity of both the near-critical and zero angles. By contrast, the stronger-contact skeleton remains stable.

% A comparison with simultaneous visualization of the surface show that for the first tilt of a fresh-prepared packing, such internal rearrangements exhibit a weak correlation with on-surface grain rearrangements. However, the latter are irreversibly aged for repeated tilts unlike the mirror-type activity in the bulk.
%
%Valid PACS numbers may be entered using the \verb+\pacs{#1}+ command.
\end{abstract}

% PACS, the Physics and Astronomy
                             % Classification Scheme.
\vspace{2pc}
\maketitle

%\begin{quotation}
%The ``lead paragraph'' is encapsulated with the \LaTeX\ 
%\verb+quotation+ environment and is formatted as a single paragraph before the first section heading. 
%(The \verb+quotation+ environment reverts to its usual meaning after the first sectioning command.) 
%Note that numbered references are allowed in the lead paragraph.
%
%The lead paragraph will only be found in an article being prepared for the journal \textit{Chaos}.
%\end{quotation}
%
\section{Introduction}\label{sec:intro}
%First-level heading:\protect\\ The line
%break was forced \lowercase{via} \textbackslash\textbackslash}
%}
A major issue of industrial and geophysical interest is the knowledge and the understanding of the stability of a granular packing. Evaluating and possibly modifying the stability of a packing is essential to predict for instance avalanches, landslides or the blocking of silos. Even when the external force applied to the packing does not induce a flow, it may drive some microdisplacements as the system adapts to the constraints. These events are precursors of the macroscopic failure; they can change the stress distribution within the packing, thus controlling its stability. The transition from a "static" state to a flowing state is challenging for condensed matter physics. This transition, which can be seen as a "jamming-unjamming transition" with astonishing analogies between granular and glassy systems, has recently triggered a wealth of publications \cite{Knight1995, Ohern2001, Hartley2003, Marty2005, Corwin2005, Xu2009}.\\
Granular system driven quasistatically  out of equilibrium exhibit memory effects and local, intermittent, rearrangements of grains \cite{Deboeuf2005, Kabla2005}.
Previous works \cite{Aguirre2000, Boltenhagen1999, Aguirre2001} show that parameters like humidity, system dimensions, friction between grains, bottom roughness, or packing fraction can influence the value of the maximum angle of stability of a packing. Aguirre et al. \cite{Aguirre2000} showed that the influence of the number of grain layers on the stability of a packing is noticeable up to about ten layers, while it becomes independent of it for larger layer numbers.
 The angle at which an avalanche starts was further found to increase with the packing fraction \cite{Aguirre2001}. Scheller et al. \cite{Scheller2006} observed an evolution of the packing fraction by successive jumps when the tilt angle of a 2D granular monolayer increases. These studies evidence clear signatures of internal reorganizations in a 2D packing and suggest the existence of ageing and memory effects during the inclination process. 
%Moreover, numerical simulations of inclined 2D grain packings, performed by Deboeuf et al. \cite{Deboeuf2005} show the existence of hysteretic behaviors during consecutive tilt cycles.    

%In this work, we demonstrate experimentally the existence of ageing and memory effects at the surface of a 3D granular packing during repeated tilt cycles. The superficial activity of the packing is monitored by classical digital imageing techniques as done in previous related experimental works on inclined granular packings\cite{bretz1, bideau1}. 
%After having described the experimental set-up and the measurement methods, we present the time evolution of the rearranged area due to instantaneous grain motions at the surface of the packing when the latter is inclined from 0 to the angle of avalanche, $\theta_{a}$. Then, we describe the evolution of the superficial activity during successive tilt cycles between 0 and a given "ageing angle" $\theta_{m} < \theta_{a}$. Finally, we depict and discuss the recovery of the activity when the inclination angle exceeds the ageing angle $\theta_{m}$.
%The stability of a granular packing %slowly driven towards its instability threshold 
%is strongly linked to
%Studies of complex and essentially independent dynamics 
%the evolution of the strong- and weak-contact networks~\cite{Radjai1998}.
% in granular materials attract much attention \cite{Radjai1998, Deboeuf2005}. 
Numerical simulations of cyclic tiltings of 2D granular piles emphasized the existence of hysteretic behaviors \cite{Deboeuf2005}. They also demonstrated the relevance of the two-phase description, in terms of weak and strong contacts~\cite{Radjai1998}, thus 
%confirming earlier 
correlating with earlier observations \cite{Zaitsev2008,Zaitsev2005} which indicate quite independent evolution of the weak- and strong-contact networks. The application of nonlinear acoustic techniques \cite{Zaitsev2008} made it possible to detect transitional modifications of the weak-contact fraction in the bulk of granular packings by observing acoustic signal components that had arisen in the material due to its nonlinearity. The key point is that such components are dominated by the weak-contact contribution \cite{Zaitsev1995,Tournat2004,Tournat2004b,Zaitsev2005}. The observation of nonlinear acoustic precursors of avalanche approaching agrees well qualitatively  with the simulated intermittent modifications of the inter-grain contact network in the bulk of tilted 2D packings \cite{Staron2002,Staron2006} and the observations of on-surface displacements \cite{Nerone2003}. 
The previous 2D simulation results were reanalyzed in detail by Henkes et al.~\cite{Henkes2010} to examine the link between the avalanche process and a global isostaticity criterion.
They emphasized  the role of clusters of particles with contacts at the Coulomb threshold, which concentrate the local failure and grow in size when approaching the avalanche.
An open  question remains as regards whether the contact rearrangements in the bulk are directly related to the on-surface ones and what the extent of independence of the weak- and strong-contact configurations is.\\
%In this work, we demonstrate experimentally the existence of ageing and memory effects at the surface of a 3D granular packing during repeated tilt cycles. The superficial activity of the packing is monitored by classical digital imageing techniques as done in previous related experimental works on inclined granular packings\cite{bretz1, bideau1}. 
%After having described the experimental set-up and the measurement methods, we present the time evolution of the rearranged area due to instantaneous grain motions at the surface of the packing when the latter is inclined from 0 to the angle of avalanche, $\theta_{a}$. Then, we describe the evolution of the superficial activity during successive tilt cycles between 0 and a given "ageing angle" $\theta_{m} < \theta_{a}$. Finally, we depict and discuss the recovery of the activity when the inclination angle exceeds the ageing angle $\theta_{m}$.
Here, we demonstrate experimentally the existence of ageing and memory effects at the surface and on the weak contact network. This is done by combining a classical digital imaging techniques~\cite{Bretz1992,Nerone2003} as well as a nonlinear acoustic method. 
In doing so, we prove that experimental results on 3D packings share common features with results of 2D simulations \cite{Deboeuf2005,Henkes2010}.\\
In section~\ref{sec:exp} we present the experimental setup and the techniques used to monitor surface activity and modifications of the weak-contact network.
Section~\ref{sec:surface} is devoted to the characterization of the surface of the packing. 
%After having described the experimental set-up and the measurement methods, 
First, we present the time evolution of the rearranged area due to instantaneous grain motions at the surface of the packing when the latter is inclined from $0^\circ$ to the angle of avalanche, $\theta_{a}$. Second, we describe the evolution of the superficial activity during successive tilt cycles between 0 and a given "ageing angle" $\theta_{m} < \theta_{a}$. We also depict and discuss the recovery of the activity when the inclination angle exceeds the ageing angle $\theta_{m}$. Third, we summarize the  results obtained and justify the use of non-linear acoustic methods to study the contact network.
In section~\ref{sec:acoustic} the acoustical measurements are reported. After having compared the information provided by the two methods, we evidence the ageing of the weak-contact network  and we describe the occurrence of "mirror-type" rearrangements. Finally, we show that the aged weak-contact configuration exhibits spontaneous restoration towards the initial state when the packing remains at rest for tens of minutes.

% Here, we apply our nonlinear acoustic technique to cyclic tilting of granular piles to 
%confirm 
%verify experimentally the results observed in the 2D simulations of Deboeuf et al \cite{Deboeuf2005}.
\section{Experimental Methods}\label{sec:exp}
%First-level heading:\protect\\ The line
%break was forced \lowercase{via} \textbackslash\textbackslash}
%}
\subsection{Experimental set-up}
\begin{figure}[htbp]
\begin{center}
\includegraphics*[width=0.6\textwidth]{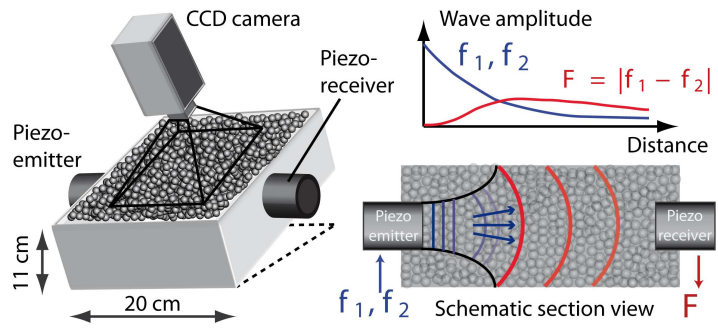}\caption{Left: Schematics of the experimental setup. Right: Acoustic probing configuration and spatial dependence of the acoustic frequency components.}%
\label{fig_setup}%
\end{center}
\end{figure}

The experimental setup consists of a slowly rotating rectangular plexiglass box containing glass beads (see figure~\ref{fig_setup}).  
%\begin{figure}[htbp!]
%\begin{center}
%		\includegraphics*[width=0.6\textwidth]{sketch} %{precursors_cambridge.jpg}
%		\caption{A sketch of the experimental set-up}
%	\label{fig:1}
%	\end{center}
%\end{figure}
%
The system is constituted of a heavy table and of a frame which can be inclined at different rates by means of a threaded stem linked to a \textcolor{black}{DC motor. Foams were used to damp vibrations from this motor.} The rotation speed can be varied from \textcolor{black}{$0.02^\circ.\mbox{ s}^{-1}$}  %$1.2^\circ /\mbox{min}$ 
 to \textcolor{black}{$(1/6)^\circ.\mbox{s}^{-1}$}. %$10^\circ /\mbox{ min}$.
 The box, which contains grains, and a camera used to scan the free surface are both fixed on the frame. The inclination angle is measured with an inclinometer fixed also on the frame. The precision of the angle measurement is $0.1^\circ$.
The size of the box is $30\times 20\times 11\ \mbox{cm}$.
We use glass beads of two types with the same diameter, $(3 \pm 0.1)\ \mbox{mm}$, similar to those used in experiments \cite{Nerone2003,Zaitsev2008} as well as smaller beads $(2 \pm 0.1)\ \mbox{mm}$ as in studies \cite{Tournat2004,Tournat2004b,Zaitsev2005}.
% and the granular packing is constituted of glass beads of diameter $3\ \pm \ 0.1\ mm$. 
In all experiments, the thickness of the packing is kept constant and equal to $8\ \mbox{cm}$. The bottom of the box is prepared for once by gluing down glass beads similar to those that constitute the piles on a flat piece of wood. The ambient humidity is kept constant at $50 \%$  to avoid effects of capillary or electrostatic forces.
To ensure reproducible results, we applied in all experiments the same procedure for the preparation of the packing.
First, a grid is put at the bottom of the box, then the box is filled with glass beads and the height of the packing is made equal to a chosen value. This is done by eroding the free surface with a bar until it reaches the chosen height marked on the lateral walls of the box. Finally, the grid is removed from the box. Glass beads cross the grid and ensure in that way the homogeneization of the packing.
Then, the height of the packing is adjusted again as mentioned before.
\textcolor{black}{The resulting initial packing fraction (evaluated by measuring the height of the packing) is $0.594\pm 0.001$.}
During an experiment, the free surface is scanned with a camera interfaced to a computer equipped with an image processing software (ImageJ). The resolution of the camera is $768\times 512$ pixels and $256$ gray levels. The acquisition speed can be chosen between $1$ and $25$ frames s$^{-1}$. 
  All results  based on image-acquisition presented in the present work were obtained with an inclination velocity of %$3^\circ$ per minute 
 \textcolor{black}{$0.05^\circ$ s$^{-1}$}  and an acquisition rate of $5$ frames s$^{-1}$. These choices are not critical, since we do not observe significant variations of our results when the inclination speed is varied in the aforementioned angular velocity interval.\\
% \subsection{Optical technique}
%\label{sec:opttech}

\subsection{Image processing} 
 The image processing is composed of three steps. First, all images of the movie recorded during the inclination process are extracted. In images, centers of grains appear as bright spots. Each image is subtracted from the next image. In that way, grains which have moved are detected. The elements of the "difference" images correspond to the positions of the centers of grains before and after the rearrangements. Then, these images are binarized and noise is reduced by applying a gray-level thresholding method. To recover the initial size of rearrangements, we apply a dilation algorithm, followed by a hole filling procedure. Finally, a particle analyzer is used to extract the position, the area, and the eccentricity of clusters of synchronously displaced grains. The conversion between pixels and glass bead diameter is obtained by calibration. The above-described image analysis method allows us to evaluate the area fraction which is rearranged between two successive images.\\

\subsection{Acoustical methods}\label{sec2C}
The acoustic method is similar to the one described in \cite{Zaitsev2008}. Acoustic pump waves composed of two frequencies $f_1$ and $f_2$ are emitted by a piezo-transducer (with an active surface radius $a=1.75 \ \mbox{cm}$) in contact with the beads at mid-depth of the granular layer (see figure~\ref{fig_setup}). A piezo-receiver placed on the opposite wall allows to detect acoustic signals at $f_1$, $f_2$, as well as some acoustic noise generated in the medium and the difference frequency signal $F=|f_1-f_2|$ arising
%present 
from the nonlinearity of the medium \cite{Zaitsev2008,Tournat2004}. The received signal is recorded using a spectrum analyzer in the form of time dependent spectra, from which the temporal evolutions of specific frequency component amplitudes can be extracted with a $39\mbox{ ms}$ time step. The emitted frequencies $f_1$ and $f_2$ (with an acoustic strain amplitude $\tilde{\varepsilon}_A\simeq 10^{-7}$) are chosen around $10\ \mbox{kHz}$ resulting from a compromise between the efficiency of the transducers and a weak acoustic scattering by the medium. $F=|f_1-f_2| \simeq 1.5\ \mbox{kHz}$ is chosen lower than $f_1$ and $f_2$ to ensure a low damping in the medium during propagation. Under these experimental conditions, the received acoustic energy is mostly coming from the waves propagating through the grains and their contacts (the solid skeleton) and not through the air.  
The presence of an in-depth acoustic velocity gradient (due to the nonlinearity of the contacts together with the increasing static pressure with depth \cite{Gusev2006}), prevents us from giving a simple picture of the acoustic propagation in this configuration. The acoustic propagation is multimodal and the several expected modes are guided by the velocity gradient \cite{Gusev2006} as well as by the layer boundaries (free boundary at the top, rigid at the bottom). From an estimation of the velocity $v\simeq 100\mbox{ m/s}$ of bulk longitudinal waves at a $4\ \mbox{cm}$ depth (static pressure $\simeq 600\mbox{ Pa}$), the wavelength of the emitted waves is $\lambda_{1,2} \simeq 1\ \mbox{cm}$. For the nonlinearly generated difference frequency $F$, the wavelength estimation gives $\Lambda \simeq 8\ \mbox{cm}$ a value comparable with the layer thickness. We notice that resonance effects between transducers are not observed in gravity stressed granular packings due to the important acoustic damping at these frequencies. The diffraction length of the emitted acoustic beam neglecting the elasticity gradient is $\ell_d = \pi a^2 /\lambda_{1,2}\simeq 10\ \mbox{cm}$. Therefore, the difference frequency wave is mainly generated from a collimated pump beam which has not yet reached the free surface of the layer. The nonlinear wave generation is thus expected to be weakly sensitive to the surface events monitored by the camera.\\
Moreover, as shown in \cite{Tournat2004,Zaitsev2008} and references therein, there are essential differences in the roles of the weakest contacts and those of the average (and strongly) loaded contacts for the propagation of linear and nonlinear acoustic waves. Let us consider a single Hertzian contact between two beads, with a static deformation $\varepsilon_0$, and subjected to a dynamic (acoustic) deformation of amplitude $\tilde{\varepsilon}_A$. A Taylor expansion of the Hertz relation \cite{Johnson1987} for $|\tilde{\varepsilon}_A|\ll |\varepsilon_0|$ provides for the dynamic perturbation, a linear elastic modulus $E^{\ell} \propto \varepsilon_0^{1/2}$ while the nonlinear quadratic modulus is $E^{n\ell} \propto \varepsilon_0^{-1/2}$ \cite{Tournat2004}. {(A typical static pressure of $600\mbox{ Pa}$ between two of our glass beads gives $\varepsilon_0 \simeq 10^{-6}$ as the average contact strain)}. Consequently, the smaller the static strain (or stress) of a contact, the higher the contact nonlinearity. 
%{\color{magenta} It is a well-know fact that in real random packings, along with average-loaded contacts (and a small portion of even stronger loaded ones) there is a significant portion of contacts whose loading is significantly smaller, down to a few percents of the average value and even less.  If the initial loading of the weak contacts is characterized by a small parameter $\mu\ll1$, the relative contribution of the weak contacts to the linear elastic modulus is $E^{\ell}_{weak}/E^{\ell} \propto (\mu\varepsilon_0)^{1/2}/\varepsilon_0^{1/2}=\mu^{1/2}\ll1$, whereas the analogous relative contribution to the nonlinear modulus $E^{\ell}_{weak}/E^{\ell} \propto (\mu\varepsilon_0)^{-1/2}/\varepsilon_0^{-1/2}=\mu^{-1/2}\gg1$. Thus the weak-contact portion gives only insignificant contribution to the macroscopic elastic modulus determining the elastic-wave propagation in the linear approximation, whereas its contribution to nonlinearity-generated signal components is strongly dominant.}
Similar reasoning holds for the hysteretic nonlinearity of sheared Hertz-Mindlin contacts. This explains the numerous experimental  {observations indicating}  the dominant role played by the weakest contacts (with $\varepsilon_0 < 10^{-7}$ or less) in the nonlinear acoustic generation process \cite{Tournat2004,Zaitsev2008,Tournat2004b,Zaitsev2005}. In particular here, the {demodulated} $F$ component is selectively sensitive to the weakest contacts in the medium, which in turn makes it sensitive to extremely weak modifications of contact forces induced by tilting. {Indeed, although the absolute displacements due to microscopic rearrangements of the grains at the averagely loaded and the weakest contacts are comparable,}  the relative force change of the averagely loaded contacts induced by tilting is much smaller, which ensures much smaller relative variations of the linear acoustic components.\\
In summary, the variations of acoustic amplitude of the nonlinearly generated $F$ component discussed in what follows reflect predominantly the variations in the weak contact configuration of the packing. The variations of the acoustic noise amplitude (typically, in the $2-3$ kHz range and  $128$ Hz bandwidth, aside of $f_1$, $f_2$, $F$ and the rotating-engine noise) come from both the surface movements of grains and from structural rearrangements in the bulk. This detected acoustic noise is at least ten times above the measurement noise.

\section{Surface measurements}\label{sec:surface}
\subsection{Dynamics of rearrangements}
\begin{figure*}[htbp]
	\begin{center}
	\includegraphics*[width=0.6\textwidth]{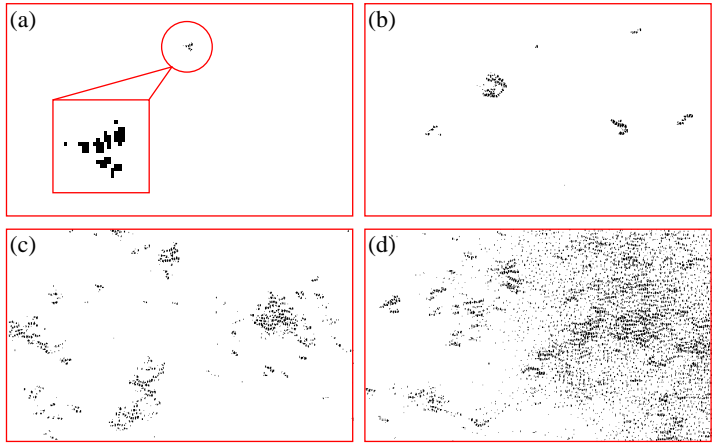}
	\caption{Examples of events observed at the surface of the packing (bead diameter $d = (3 \pm 0.1)\ \mbox{mm}$) for different inclination angles. (a) $3^\circ$ (b) $12^\circ$ (c) $20^\circ$ (d) $26^\circ$ 
	Each subfigure makes visible all rearrangements which occur between two successive images.}
	\label{fig:descriptionevenements}
	\end{center}
\end{figure*}

\begin{figure}[htbp]
	\centering
		\includegraphics*[width=0.6\textwidth]{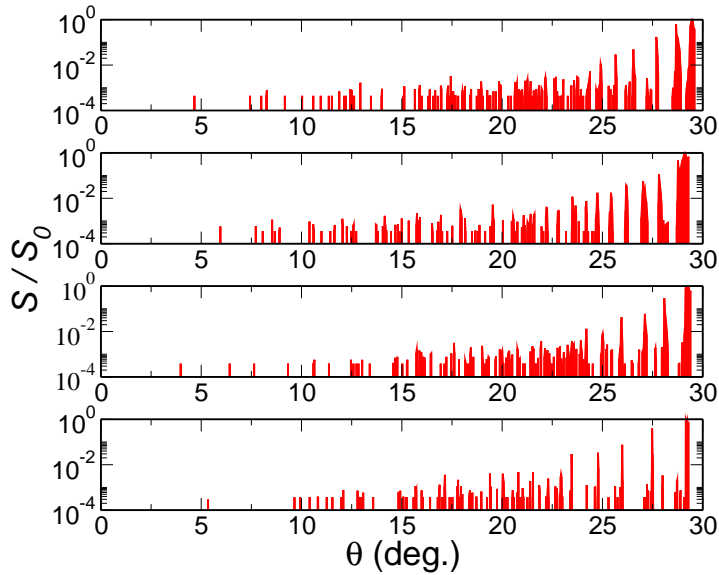}
	\caption{Evolution of the rearranged fraction of the surface with the inclination angle for different experiments carried out in the same conditions (bead diameter $d = (3 \pm 0.1)\ \mbox{mm}$).}
	\label{fig:repro_billesmates}
\end{figure}

We describe below experimental results concerning the dynamics of superficial rearrangements during a single inclination of the packing from zero angle until the avalanche onset at $\theta_{a}$. Figure~\ref{fig:descriptionevenements} shows examples of events taking place at the surface of the packing during the inclination from $0^\circ$ to $\theta_{a}$. Each subfigure shows all rearrangements which occur between two consecutive images around the  inclination angle considered. The number and size of events %are seen to 
increase with the inclination angle.
While rearrangements are isolated from each other for small inclination angles, those which take place for angles larger than $20^\circ$ involve a large fraction of the grains located at the surface of the packing. Image analysis techniques, described in the previous section, yield the surface fraction which has evolved between two consecutive images. Figure~\ref{fig:repro_billesmates} shows the evolution of the rearranged surface fraction $S/S_{0}$ with the angle of inclination of the packing for different experiments performed in similar conditions. Here, $S_0$ is the total area of the surface of the packing. %Results are seen to be reproducible. 
The rearranged fraction exhibits an overall increase with the inclination angle in a reproducible way.
A transition to an intermittent regime, in which large events occur almost periodically, takes place at about $20^\circ$. The latter events correspond to the large events named "avalanche precursors" by Nerone et al. \cite{Nerone2003}. They are due to the simultaneous movement of a large fraction of the grains of the surface of the packing in agreement with the findings of Nerone et al.~\cite{Nerone2003} and of Zaitsev et al. \cite{Zaitsev2008}. These authors report indeed that similar large events occur at angles smaller, by a few degrees, than the angle of avalanche. It is worth noticing that similar intermittent events are observed in numerical simulations of inclined 2D packings \cite{Staron2006}. The results of previous non-linear acoustic studies of Zaitsev et al. \cite{Zaitsev2008}, those discussed in section ~\ref{sec:acoustic} and those of 2D numerical simulations of Staron et al. \cite{Staron2006} all suggest that these large events are linked to intermittent reorganizations of the network of weak contacts in the bulk of the packing.
 
\begin{figure}[htbp]
	\centering
	\includegraphics*[width=0.6\textwidth]{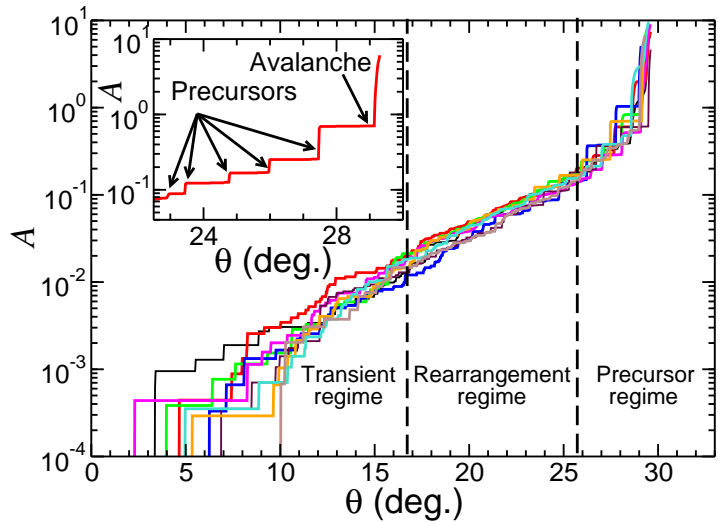} %{cumsumbillesmates2.jpg}
	\caption{Cumulated sum $A(\theta)$ (bead diameter $d = (3 \pm 0.1)\ \mbox{mm}$) for experiments performed in similar conditions. Inset : more detailed record for the last degrees before the  avalanche.}\label{fig:cumsumbillesmates}
\end{figure}

To characterize the dynamics and quantify the increase of the rearranged surface fraction during inclination, we show in figure~\ref{fig:cumsumbillesmates}, the activity of the packing, $A(\theta)$, defined as the cumulated sum over the rearranged surface normalized by $S_0$.

\begin{equation}
A(\theta)= \sum_{j=0}^{\left\lfloor \frac{\theta}{d\theta}\right\rfloor} \frac{S(jd\theta)}{S_{0}}. 
	\label{eq1}
\end{equation}

In equation (\ref{eq1}), $\theta$ is the angle of inclination of the packing and d$\theta$ is the angular interval between two consecutive images.
   
Different results found for experiments performed in similar conditions are plotted in figure~\ref{fig:cumsumbillesmates} to emphasize their reproducibility. Three regimes are evidenced in the evolution of the activity of the packing during inclination. First, for small angles, a transient regime depends significantly on the initial preparation of the packing, where some beads initially in metastable positions on the surface move. Second, a growth
regime is characterized by an exponential increase of the activity where rearrangements involving groups of several beads occur. Third, a regime dominated by the occurence of precursors, with pseudo-periodic large events, appears close to the critical angle. Last, the system avalanches. %Figure 4 strengthens the previous conclusion about the reproducibility of the behavior of the surface activity during a single tilting from 0 to $\theta_{a}$
\textcolor{black}{The most intense precursors occur for angles larger than $25^\circ$ (figure~\ref{fig:cumsumbillesmates}, rightmost dotted line), but the first, less intense, ones are more difficult to detect. The minimum value of the angle at which they can be detected, is of the order of $20^\circ$ and varies from experiment to experiment (see inset of figure~\ref{fig:cumsumbillesmates}).
For simplicity, the precursor dominated regime will be named hereafter "precursor regime".}

\subsection{Evidence of ageing}~\label{sec:ageing}        
To study the influence of the system history %of the system 
on the activity of the packing, we applied a series of consecutive forth-and-back tilting cycles to the container. During a cycle, the box is first inclined from the zero angle to an angle $\theta_{m}$ chosen to be close to the critical angle and then it is inclined back to the zero angle. The maximum inclination angle $\theta_{m}$ will be named hereafter the "ageing angle". The activity at the surface of the packing is followed during the part of each cycle with forward tilting.
  \begin{figure}[htbp]
	\centering
	\includegraphics*[width=0.6\textwidth]{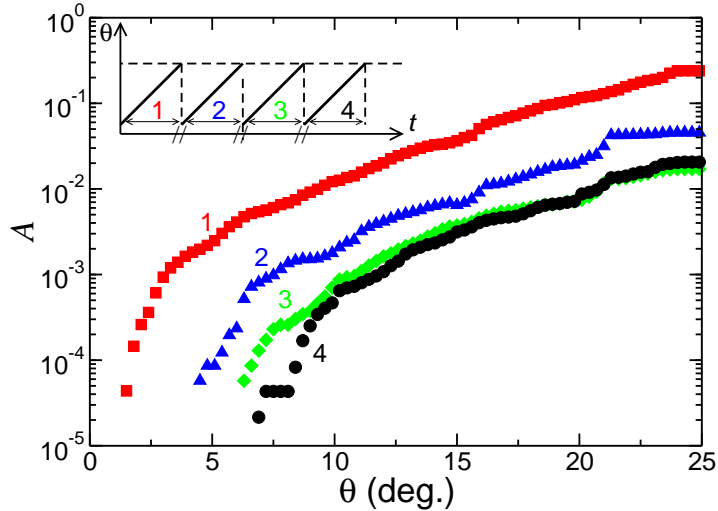}%{relaxationdynamique2.jpg}
	\caption{Ageing during several consecutive inclinations of the pile with $\theta_{m}=25^\circ$ (bead diameter $d = (3 \pm 0.1)\ \mbox{mm}$). Each curve is a mean over five independent experiments.}
	\label{fig:relaxationdynamique}
\end{figure}
   \textcolor{black}{During a given inclination, a regular overall increase of the activity of the packing with $\theta$ occurs up to $\theta_{m}$. Figure~\ref{fig:relaxationdynamique} clearly shows that the system is in a gradually evolving non-stationary state, with an overall activity decrease, during the first three cycles. After four cycles, the activity of the system reaches a limit state corresponding  to strongly reduced intensity. The surface is then in a rather stable, "aged" configuration. The system keeps  its "aged" configuration when it is inclined between 0 and $\theta_{m}$. This evolution of the activity of the packing is a clear signature of the existence of the influence of the past history of the system on its future evolution. That phenomenon, which was reported for instance in experiments of granular compaction \cite{Josserand2000}, is typical of evolution of out-of-equilibrium glassy systems. Thus, the activity of a slowly tilted granular packing depends not only on geometrical properties of the packing: bead diameter and box dimensions, initial packing fraction, number of layers, but also on the mechanical perturbations and loading applied to it before it is inclined. Using two-dimensional contact dynamics simulations of packings of discs inclined forth and back during successive cycles, Deboeuf et al. \cite{Deboeuf2005} showed that a 2D system driven in that way exhibits an ageing phenomenon related to an anisotropy of the contact orientations. They further show that the latter anisotropy is concentrated in the network of weak contacts. The observed behavior dwells in an asymmetry in the activity of the packing when it is tilted forth and back. The evolution of the contact orientations during the inclination process is irreversible.
 Contacts tend to orientate in a favored direction which depends on that of gravity.
 In any case, even if there exists a priori differences between the behavior of 2D and 3D inclined packings, our experiments on inclined 3D packings and numerical simulations of inclined 2D packings of Deboeuf et al. \cite{Deboeuf2005} exhibit similar ageing phenomena. The question of the existence of such an ageing in the weak-contact network will be addressed in section~\ref{sec:acoustic}.}

%To investigate the influence of $\theta_{m}$ on the dynamics of the system, we performed a series of four cycles as described above. The fifth and last cycle differs from the previous ones in that the system was inclined up to the angle of avalanche. Figure~\ref{fig:reprise} presents the evolution of the activity of the packing during the four first passages up to $\theta_{m}= 25^\circ$ and the activity during the final inclination up to the angle of avalanche.  

%\begin{figure}[htbp]
%	\centering
%	\includegraphics*[width=0.6\textwidth]{ageing_surface_aval} %{vieillissementmates2.jpg}
%	\caption{(Color online) Activity of the packing during the four first passages and during the final passage until the system avalanches.}
%	\label{fig:reprise}
%\end{figure}
%
%After four forth and back tiltings, the system has reached a stationary state with a limited activity (see too figure 5). When the pile is periodically inclined between 0∞ and $\theta_{m}$, the on-surface rearrangements exhibit pronounced ageing: during consecutive tiltings, the activity decreases compared to that of the first cycle.
% 
%This figure %~\ref{fig:reprise} 
%shows that the packing activity strongly increases exactly as soon as the inclination angle exceeds $\theta_{m}$. 
 
\begin{figure}[htbp]
	\centering
		\includegraphics*[width=0.6\textwidth]{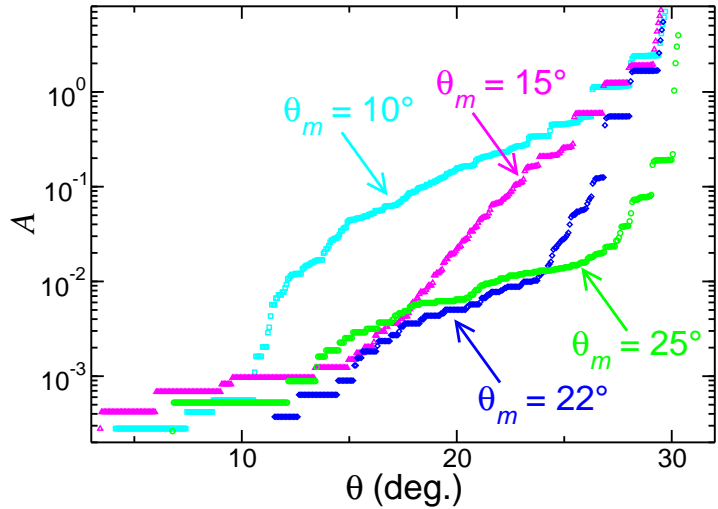} %{reprise-mates.jpg}
	\caption{Resurgence of the activity of the packing for various ageing angles $\theta_{m}= 10^\circ, 15^\circ, 22^\circ, 25^\circ$. The bead diameter is $d = (3 \pm 0.1)\ \mbox{mm}$.}
	\label{fig:reprise-mates}
\end{figure}
%The influence of the ageing angle $\theta_{m}$ on the resurgence of the activity of the packing was further investigated.
 \textcolor{black}{To investigate the influence of the ageing angle $\theta_{m}$ on the dynamics of the system, we performed a series of four cycles as described above. The fifth and last cycle differed from the previous ones in that the system was inclined up to the avalanche angle. Figure~\ref{fig:reprise-mates} shows the superficial activity of the packing for various values of the ageing angle $\theta_{m}$ during the last passage. The activity resumes precisely at the point where the inclination angle exceeds $\theta_{m}$ for any value of $\theta_{m}$ ranging between $0$ and $\theta_{a}$. The system keeps in its configuration the memory of the value $\theta_{m}$ through irreversible reorganizations during cycles in the interval 
 [0; $\theta_{m}$]. The previous observations are in good agreement with the results obtained on granular packings in a slowly rotating drum \cite{Kabla2005}. 
%When $\theta_{m}$ takes a value in the precursor regime, precursors appear only for the first inclination. 
No precursors are observed in the aged configuration (i.e. for $\theta$ lower than $\theta_{m}$). However precursors reappear for inclination angles larger than $\theta_{m}$. Figure~\ref{fig:reprise-mates} shows that precursors occur at angles $\theta$ lower than $25^\circ$ for $\theta_{m}$ less than $25^\circ$. By contrast precursors are not evidenced for $\theta$ lower than $25^\circ$ when $\theta_{m} = 25^\circ$ while they would be expected to appear as they do for the other ageing angles. The disappearance of precursors during the ageing process in the interval [ 0; $\theta_{m}$](figure~\ref{fig:reprise-mates}) highlights the irreversible nature of precursors.} 

The path to the avalanche of a given packing depends thus on the value of $\theta_{m}$ and shows a hysteretic behavior. The resurgence of the activity shown by figure~\ref{fig:reprise-mates} emphasizes once again the irreversible nature of the reorganizations in an inclined packing.
\subsection{Summary}
 The aforementioned experimental results constitute evidence of the existence of ageing and memory effects in 3D packings slowly driven to their maximum angle of stability. During inclination cycles, the activity of a granular packing decreases and reaches a stationary state.  
  %This process may be linked with the process of compaction observed in granular matter under vertical tapping \cite{Ribiere2007} where relaxations are observed to occur on long time scales and where individual motion of grains has a strong importance on the the dynamics of the system~\cite{Ribiere2005b}. 
 %ageing effects stabilize the packing. They emphasize the major role played by the orientation of gravity on the reorganization of contact networks. In particular, the activity of the packing was shown to depend strongly on the past of the system. This ageing effect has to be considered in parallel with the influence of the initial preparation of the packing (for example the initial packing fraction). These results suggest that information on the history is, in some way, stored in a specific organization of contact between grains. 
%  
%   Our experiments suggest that 
%Therefore, the non-stationary dynamics of an inclined grain packing shares relaxation properties and memory effects in common with granular media subjected to different kinds of mechanical perturbations (for instance, tapping \cite{Josserand2000}).\\

These results deal  exclusively with observations performed at the surface of the packing.
   \textcolor{black}{Numerical simulations~\cite{Staron2002,Staron2006,Henkes2010} 
 show that approaching avalanche-type instabilities are linked to the nature of the grain-grain contacts in the bulk of the material.} 
  % precursors are linked to the nature of the grain-grain contact in the bulk of the material. 
To test how relevant the latter results are for real 3D packings, it is necessary to explore the nature of the contacts within the granular medium. Imaging methods, for instance X-ray tomography~\cite{Richard2003,Aste2005}, are unable to characterize the actual nature of a contact (two spheres can be infinitesimally close without touching each other). Photoelasticity allows to visualize the weak and strong force networks in 2D systems but the application of this method to 3D systems is presently beyond reach. To the best of our knowledge the only technique that is able to probe  weak contacts (with the average loading of a few percents of the mean value) within the bulk of a granular medium and with a high temporal resolution is the non-linear acoustic method described in~\cite{Zaitsev2008}.     
In the next section we will use the latter method to show that similar effects occur in the bulk of the packing. This will allow us to compare the respective roles played by the weak and strong contact networks in the 
   %relaxation process 
  dynamics and consequently their roles in the unjamming transition. 
 \section{Acoustical measurements}\label{sec:acoustic}
 {\subsection{Pre-avalanche variations of the nonlinear signal component}}
  
{Figure~\ref{fig:comp-lin-nlin} shows an example of the acoustical record obtained during a tilting of the packing for the last $8^\circ$  before the critical angle. The noise of the avalanche itself which started at $28^{\circ}$ is not shown here. The levels of two received frequency components are presented as a function of the tilt angle: the nonlinear demodulated $F$ component and the component $f_1$ initially excited in the medium. As discussed earlier, in section~\ref{sec2C}, this record illustrates the higher sensitivity of the nonlinear $F$ component, which is preferentially influenced by the contribution of the weakest contacts in the medium (with a static strain of a few percents of the average contact static strain). Quasi-periodical variations (in the form of abrupt drops) of the nonlinear $F$ component amplitude can be easily identified in figure~\ref{fig:comp-lin-nlin} for the last $\sim 4^\circ$ before the avalanche. This qualitative trend has been analyzed in \cite{Zaitsev2008}. In what follows, in view of the higher sensitivity of the nonlinear signal $F$ component compared to the linear one, we focus on the nonlinear $F$ component analysis and in some cases on the acoustic noise as well.
The abrupt drops in the amplitude of the nonlinear $F$ component play a role analogous to the intensity bursts of the on-surface rearrangements (compare figures~\ref{fig:repro_billesmates} and \ref{fig:comp-lin-nlin}). 
%{\color{magenta} Figure~\ref{fig:comp-lin-nlin} shows an example of the variability of the demodulated $F$ component of the acoustic signal arisen due to the material nonlinearity and the primary component of the sounded signal, i.e. the component propagated in the conventional  linear sense. The records corresponds to the last $8^{0}$ before the critical angle. The noise of the avalanche itself (which started at $28^{0}$) is not shown. This record illustrates the higher sensitivity of the nonlinear components of the signal (here the demodulated $F$ component) dominated by the contribution of the weakest contacts. It should be emphasized that their initial static strain is on the order of a few percents compared to that of the average loaded contact fraction which forms the skeleton of the pile and mechanically supports it. The pronounced quasi-periodical variations (abrupt drops and returns) in the nonlinear-signal amplitude are clearly seen in Fig.~\ref{fig:comp-lin-nlin} for last several degrees before the avalanche. Other examples with the same qualitative trend can be found in \cite{Zaitsev2008}. In what follows, in view of the higher sensitivity of the nonlinear signal component compared with the linear one, we focus on the nonlinear signal and in some cases on the acoustic noise as well. 
\begin{figure}[htbp]
\center
\includegraphics*[width=0.6\textwidth]{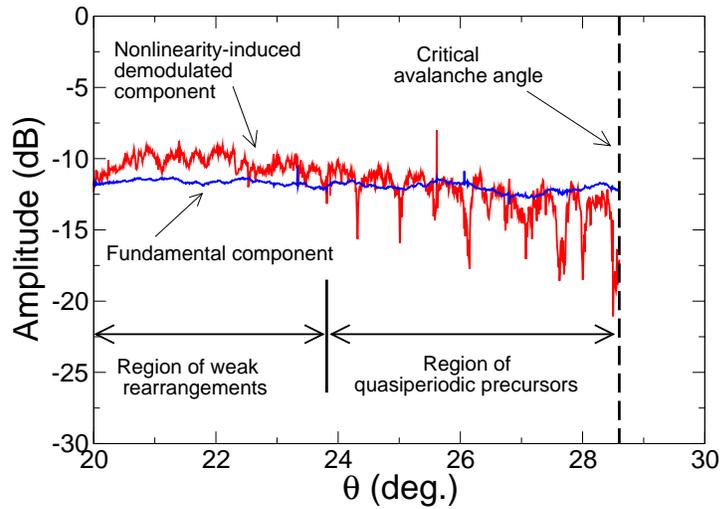}\caption{Simultaneously recorded amplitudes of the linear $f_1$ component initially radiated by the acoustic source and of the $F$ component (nonlinearly demodulated in the material). The latter exhibits stronger variations in the inclined pile for the angular range $20-28^{\circ}$. The bead diameter is $d = (3 \pm 0.1)\ \mbox{mm}$. 
}%
\label{fig:comp-lin-nlin}%
\end{figure} 
%By analogy with Fig.~\ref{fig:repro_billesmates} presenting the intensity bursts of the on-surface rearrangements of the grains, Fig.~\ref{fig:series-nonlin} shows records of nonlinear acoustic $F$ component level obtained for three different types of beads ($2 \pm 0.1\ mm$ in diameter beads and $3 \pm 0.1\ mm$ beads with two different surface properties). 
Like what was done for surface precursors (end of section~\ref{sec:ageing}), we will name hereafter "precursor region" the angular range in which the most intense, quasi-periodical, acoustical drops occur.
% and \ref{fig:series-nonlin}). 

In the general case, the level of the demodulated signal is determined by many factors, among which are the number of the weakest contacts and their loading, the amplitude and attenuation in space of the initially excited $f_1$ and $f_2$ components, \ldots. Due to the weak amplitude variations of $f_1$ and $f_2$, the dominant contribution to the variation of the $F$ component amplitude is expected to arise from the variations of the nonlinear elastic properties of the packing, i.e. the weak-contact properties. Numerical simulations \cite{Staron2002,Staron2006,Henkes2010} report the existence of a temporarily unloading, up to complete break, of the "mobilized" weakest contacts during microrearrangements. This observation is consistent with the modeling of the demodulated-signal amplitude of \cite{Zaitsev2008} which shows that drops in the nonlinear signal amplitude are produced when the number of weak contacts contributing to the nonlinear acoustic process is temporary diminished. Consequently, abrupt drops in the nonlinear $F$ component amplitude can be interpreted as a temporary decrease in the number of the weakest contacts during the tilt-induced small rearrangements. It should be recalled that to our knowledge, no other experimental technique is able to probe in real time such fast transient variations in the state of the weakest contacts:  millisecond time scales (e.g., the time step in figure~\ref{fig:comp-lin-nlin} is $39$~ms) and subnanometer scale acoustic displacements.

%    
%\begin{figure}[htbp]
%\center
%\includegraphics*[width=0.6\textwidth]{series-nonlin_new}\caption{(Color online)  Examples of the evolution %of the nonlinear $F$ component amplitude %recorded 
%over the entire angular range from $0$ degree to the critical angle. For a better visual comparison with Fig.~\ref{fig:repro_billesmates}, the direction of the vertical axis is reversed. Thin solid lines show the slow trends averaged over the window of $1 /20$ of the entire record, which were subtracted when finding the Hurst parameter. Curve marked $L$ in panel  (a) shows for comparison the weak variability of the fundamental (linearly propagated) signal component. 
%}%
%\label{fig:series-nonlin}%
%\end{figure}

%Returning to the discussion of the features of the nonlinear-signal variations, one can  see from the examples presented in Fig.\,\ref{fig:series-nonlin} a remarkable 
 
We now analyze the transition from a noisy-like behavior to a quasi-periodic appearance of drops in the $F$ component amplitude (like in the example shown in figure~\ref{fig:comp-lin-nlin}%\ref{fig:series-nonlin}
) as a function of the tilt angle. %In Fig.\,\ref{fig:series-nonlin}, although all the signals exhibit the same trend of variability increase with the tilt angle, it is difficult to average them directly in order to extract a general quantitative behavior. By analogy 
The acoustic signals in different experimental runs exhibit similar trends of variability increase with the tilt angle like in figure~\ref{fig:comp-lin-nlin}. Nevertheless the exact positions of the extrema  certainly do not coincide  (like for the visual records in figure~\ref{fig:repro_billesmates}). Therefore, it is difficult to directly average such signals  in order to extract a general quantitative behavior. In view of this, by analogy 
with the cumulative characterization of the on-surface activity in the previous section, we characterize statistically the $F$ component amplitude variations with the help of the Hurst parameter \cite{Hurst1951}. \textcolor{black}{We recall that for a discrete signal $Y(u)$ (with $1\leq u \leq N$), the Hurst parameter $H(j)$ %for a discrete-signal record $Y(i)$ with the length $1\leq i \leq N$ 
within a portion $1 < j < N$ can be expressed as the ratio between the accumulated peak-to-peak variation $R(j)$ and the standard deviation $S(j)$ % within a portion $1<j \leq N$ of the entire record 
\cite{Embrechecht2002}:}
\begin{equation}
H(j)=R(j)/S(j).
\label{eqh1}
\end{equation} 
The accumulated peak-to-peak variation $R(j)$ is given by
\begin{equation}
R(j)=\max_{1\leq i \leq j}(X_{i,j})-\min_{1\leq i \leq j}(X_{i,j}), 
	\label{eqh2}
\end{equation}
\begin{equation}
\mbox{with }X_{i,j}= \sum_{u=1}^{u=i\leq j} [Y(u)-M_{j}], \mbox{ and }M_{j}= \frac{1}{j}\sum_{u=1}^{j} Y(u).
\label{eqh2bis}
\end{equation}
The standard deviation $S(j)$ is defined by 
\begin{equation}
S_{j}= \Bigg\{\frac{1}{j}\sum_{i=1}^{j} [Y(i)-M_{j}]^2\Bigg\}^{1/2} .
\label{eqh3}
\end{equation}
Then the logarithm $\log [H(j)]$ is plotted against $\log(j)$ and the slope of the dependence is determined. This slope is called the Hurst exponent. For a random signal without a trend its value equals to $1/2$ (i.e. the normalized peak-to-peak variation increases as a square root of the signal length). If a deterministic trend is present in the signal, the Hurst exponent value becomes close to unity (and such signal is called persistent). A Hurst exponent smaller than $1/2$ corresponds to so-called antipersistent signals, for which an increase is likely to be followed by a decrease.
%Looking at the precursor region with abrupt variations like in Figs.\,\ref{fig:comp-lin-nlin} and \ref{fig:series-nonlin}, both the peak-to-peak variation and standard deviation almost stabilized in this region, so that the Hurst exponent should also exhibit a pronounced trend to stabilization. Therefore, the plots of the normalized Hurst parameter should give a much stabler representation of the processes, even if there is no much sense to directly average the records similar to those in Figs.\,\ref{fig:comp-lin-nlin} and \ref{fig:series-nonlin} (which can only smear the abrupt drops-precursors with random positions). 
The cumulated representation of the Hurst parameter allows to compare different records such as that of %those of Fig.\,\ref{fig:series-nonlin} 
figure~\ref{fig:comp-lin-nlin} as well as to distinguish the precursor region from the random noisy-like region. 

\begin{figure}[htbp]
\center
\includegraphics*[width=0.6\textwidth]{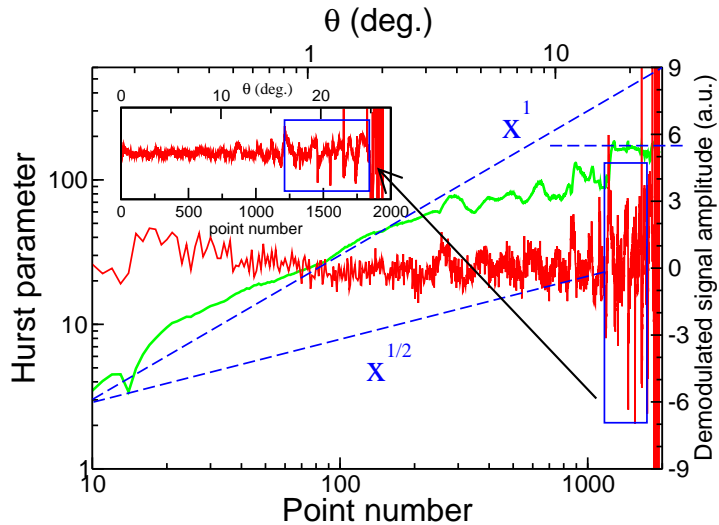}\caption{$F$ component acoustic amplitude recorded during tilting up to the critical angle superposed with the corresponding Hurst parameter. The bead diameter is  $d = (3 \pm 0.1)\ \mbox{ mm}$. The inset shows a zoom on a linear scale of the precursor region for which the Hurst parameter stabilizes. The characteristic slopes $x^{1}$, $x^0$ and $x^{1/2}$ are shown by dashed lines.} 
\label{fig:hurst-single}
\end{figure}

\begin{figure}[htbp]
\center
\includegraphics*[width=0.6\textwidth]{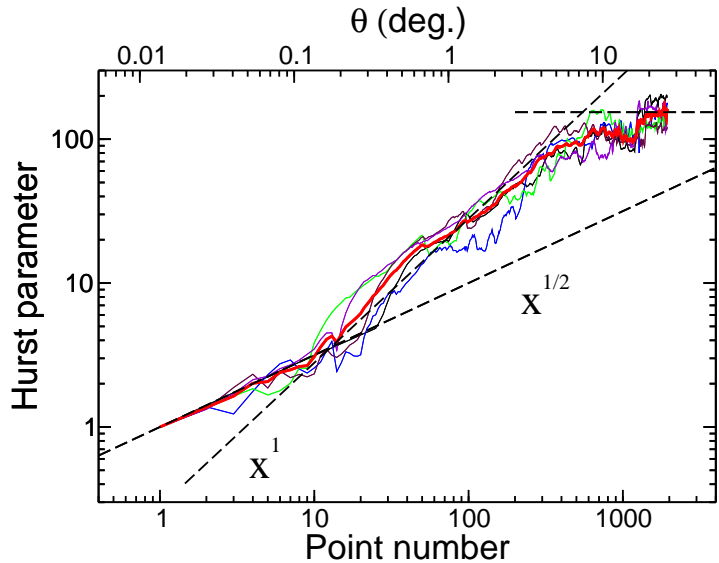} %{hurst-aver}
\caption{Evolution of the Hurst parameter during the tilting of the packing (bead diameter $d = (3 \pm 0.1)\ \mbox{mm}$) from zero to the critical angle. The thick curve shows the average over five records, similar to that in %those in
 figure~\ref{fig:comp-lin-nlin} but for the same beads. Individual dependences are shown by thin lines. The Hurst parameter stabilizes close to the critical angle. This behavior is observed for the averaged and the individual records. The dashed lines show the characteristic slopes $x^{1}$, $x^{1/2}$, and $x^{0}$.} 
	\label{fig:hurst-aver}%
\end{figure}

Figure~\ref{fig:hurst-single} shows the Hurst parameter for a record similar to the one reported in figure~\ref{fig:comp-lin-nlin}.
%another record of the same type as in Fig. 7.
%one of the records of Fig.\,\ref{fig:series-nonlin}. 
Since the slow signal changes are mostly associated with the slow evolution in the loading of the contacts during the titling, we subtracted the slow trend %obtained 
by averaging over sliding windows of $1/20$ of the entire length of the record and then applied the above-described procedure for determining the Hurst parameter. The Hurst parameter stabilizes (a bright antipersistent behavior) in the region of the quasi-periodic drops in the $F$ component amplitude (figure~\ref{fig:hurst-single}). In the initial stage, before the precursors, the signal behavior is closer to a persistent-type behavior with a Hurst exponent close to unity. The thick curve in figure~\ref{fig:hurst-aver} shows the Hurst parameter averaged over five different records (shown by thin lines). The average curve and the individual ones are characterized by a Hurst exponent close to unity in the region before the precursors and characterized by a constant Hurst parameter in the precursor region. Consequently, such a cumulative characteristic appears to be well reproducible over different experimental runs, in contrast to the particular drop features. This constitutes a useful way to determine the onset of precursors and the approaching of the stability loss of the packing.}

 \subsection{Comparative studies of the acoustic signal from the material bulk and of surface rearrangements of the grains} 
To study the relation between the surface activity and the contact rearrangements in the bulk, simultaneous visual and acoustic  measurements were performed both during the ascending and the descending phases. 
%Two types of packing were used:  fresh-prepared ones and packings subjected to several cycles of tilting. For the latter, each time the tilting was stopped at about $26^{0}$, just before the critical angle of $\sim27^{0}..28^{0}$. The visual observations confirmed the intuitive expectation that the most intensive activity occurred for fresh-prepared packings, on the surface of which, even after mechanical smoothing,  a significant  number of metastable easily perturbed grains remained. 
%Both the nonlinear acoustic signal variations and surface activity are qualitatively consistent with earlier reported visual data \cite{Nerone2003}, nonlinear acoustic observations \cite{Zaitsev2008} and numerical simulations \cite{Staron2002,Staron2006} according to which the intensity of the precursors increases closer to the avalanche.  

As reported in section~\ref{sec:surface}, in earlier experimental data~\cite{Nerone2003}, in nonlinear acoustic observations \cite{Zaitsev2008} and in numerical simulations \cite{Staron2002,Staron2006} the intensity of the precursors increases when the packing is inclined closer and closer to the angle of avalanche.
\begin{figure}[htbp]
\begin{center}
\includegraphics*[width=0.6\textwidth]{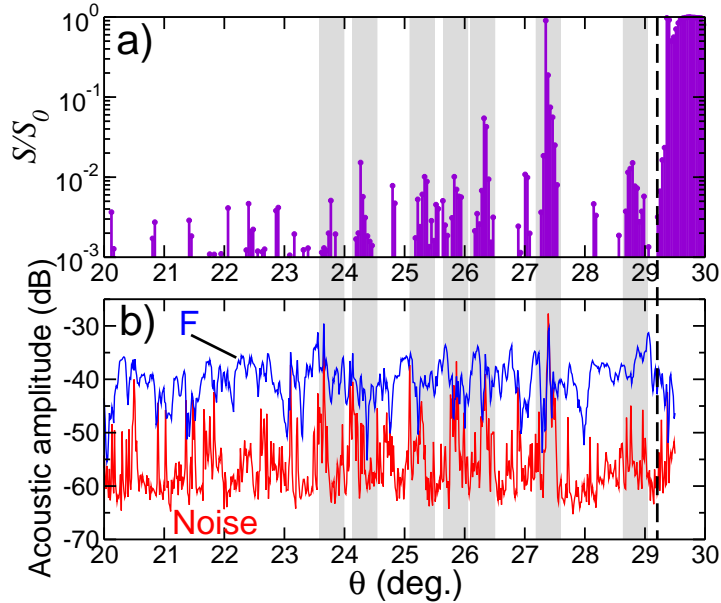} %{figvid2.pdf}
\caption{Simultaneously recorded rearranged fraction of the surface $S/S_0$ (a), and of $F$ variations and acoustic noise around $2816$ Hz (b), in the inclined pile for the angular range  $20-27^\circ$ (bead diameter $d = (3 \pm 0.1)\ \mbox{mm}$). 
%The intensity $S/S_{a}$ of the on-surface displacements is normalized to the value occurred for the  resultant avalanche. 
The gray regions emphasize a high-degree of  co-occurrence of variations of the superficial activity, of those of $F$ due to weak-contact rearrangements and that of noise bursts. }%
\label{Vis1}%
\end{center}
\end{figure}
These precursors exhibit a quasi-periodic behaviour and become less frequent than the initial small-amplitude rearrangements. An example of the dependence of the normalized surface activity intensity $S/S_0$ with the tilt angle of the pile is shown in figure~\ref{Vis1}(a) while the simultaneously recorded nonlinear $F$ signal and acoustic noise are shown in figure~\ref{Vis1}(b). 
The figure shows only the range of large angles  ($20^{\circ}-27^{\circ}$), where the surface activity is significant. For some  peaks of the surface activity (marked by gray regions in figure~\ref{Vis1}), there are simultaneous rapid variations of the $F$ signal related to the weak-contact rearrangements in the material bulk. However, such coincidences are not observed for all peaks, a result which makes the relation of the surface events to the weak-contact rearrangements in the bulk not so obvious. In addition, a significant number of expected correspondences do exist between the on-surface activity and the recorded noise bursts in the tilted pile. \textcolor{black}{We evaluated the correlation function between the peaks in the on-surface activity and the absolute value of the derivative of $F$ with respect to the angle. A rough estimation based solely on three experiments, shows that, for simultaneous records of the two signals, the cross-correlation displays a maximum value of typically  $0.25$ when the delay time is zero. The correlation-peak amplitude for the visual and acoustic signals recorded for the same tilt appeared to be noticeably greater ($\approx 1.5$ times) than the amplitude obtained by correlating a visual signal taken from one tilt and an acoustic signal taken from another, independent tilt. This is an indication that the correlation is weak but discernible. Two facts can explain this weak correlation. First, a large contact reorganisation in the volume (detected by the acoustic method) does not necessary bring out grain motion at the surface (detected by the optical method). Second, large grain displacements at the surface are related to a reorganisation of contacts at the surface which can be more or less correlated to the reorganization of contacts in the volume.}

\begin{figure}[htbp]
\center
\includegraphics*[width=0.6\textwidth]{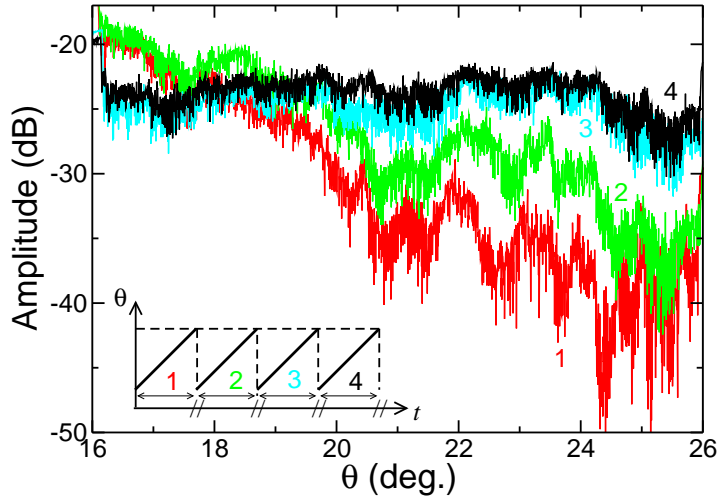}%{ageing_gradual.pdf}
\caption{Example of the weak-contact network ageing during several consecutive tilts of the pile as evidenced by the nonlinear $F$ signals shown here for the last $10^\circ$ before $\theta_m=26^\circ$ (bead diameter $d = (2 \pm 0.1)\ \mbox{mm}$).}%
\label{ageing}%
\end{figure} 
\subsection{ageing of the weak-contact rearrangements in the bulk} 
As described in section~\ref{sec:surface}, the surface activity exhibits strong ageing after a few consecutive cycles of forth-and-back tilting whith a maximum tilting angle $\theta_m$, smaller than the critical angle of value $\sim27^\circ-28^\circ$ . 
%For the surface rearrangements, it is physically clear that the surface grain displacements result in a more stable, smoothed surface profile. Therefore, for periodical tilting between zero and nearly critical angle, the surface activity exhibits strong ageing (see section~\ref{sec:surface}). We apply a series of consecutive forth-and-back tilting cycles where tilting was stopped at about $\theta_m = 26^\circ$, just before the critical angle of $\sim27^\circ-28^\circ$. The visual observations and image analysis confirmed that the most intensive activity occurred for fresh prepared packings, on the surface of which, even after mechanical smoothing, a significant number of metastable easily perturbed grains remain. For subsequent cycles, the surface activity is reduced by orders of magnitude compared to the first cycle (see section~\ref{sec:surface}).  
These observations coincide with those of numerical simulations of 2D inclined packings \cite{Deboeuf2005,Henkes2010}. Deboeuf et al. \cite{Deboeuf2005} relate this "ageing" phenomenon to an anisotropy of the contact orientations, concentrated in the network of weak contacts. Even if 2D and 3D systems differ in many respects, it seems reasonable to expect similar behaviors for the bulk rearrangements of the least stable weak-contact subnetworks.
%The observations have shown that such expectations are valid only to a certain extent.  
Figure~\ref{ageing} shows  several such consecutive records of the variability of the weak-contact subnetwork, observed via the nonlinearity produced $F$ signal for $\theta_m = 26^\circ$ in the angular range in which $F$ varies the most \cite{Zaitsev2008}. The overall amplitude of variation of $F$ is seen to decrease significantly cycle after cycle. Four consequent tilts were sufficient for the weak-contact network to reach an aged configuration, for which no appreciable further variations were noticeable in the angular range studied.
%\begin{figure}[ptb]
%\center
%\includegraphics[height=5cm]{fig_steps2.pdf}\caption{Schematically shown overlapped angular ranges of tilt angles in the experiments on ageing/reactivation of weak-contact rearrangements.}%
%\label{fig_steps}%
%\end{figure} 
%
 \begin{figure}[htbp]
\center
\includegraphics*[width=0.6\textwidth]{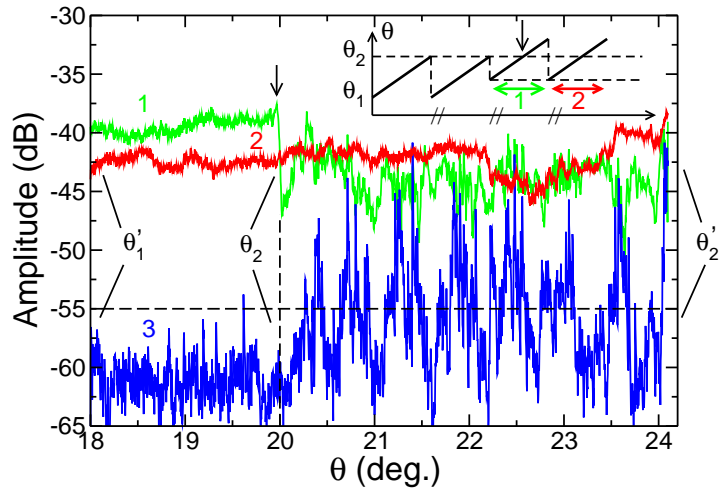} %{reactiv.pdf}
\caption{Example of reactivation of the weak-contact network rearrangements upon surpassing the previous upper angle $\theta_2=20^\circ$ (curve 1). The noise level (curve 2) is also sensitive to the reactivation. The next upward tilt again exhibits an ageing stage (curve 3). The bead diameter is $d = (2 \pm 0.1)\ \mbox{mm}$.}%
\label{reactiv}%
\end{figure} 

Although four tilts were also found for the ageing of the surface activity (see section~\ref{sec:surface}),
it is necessary to wonder whether the transition to the new more stable state could be caused by the acoustic field itself rather than by the variation in the gravity-field orientation. In order to clarify this, we performed several consecutive tilts of a freshly prepared packing with a several-degree excursion up to a fixed threshold angle below the critical one. After several tilts (for example, in the range $\theta_1=14^{\circ}\leq \theta \leq \theta_2=20^{\circ}$) the variations in the acoustic signal are strongly reduced like in figure~\ref{ageing}. Then the packing is tilted to a larger angle (in the case discussed, the next range is $\theta_1'=18^{\circ}\leq \theta \leq \theta_2'=24^{\circ}$; see the insert in Figure~\ref{reactiv}). If the ageing of the weak-contact network resulted from the acoustic perturbation, then the increase in the tilt angle above $\theta_2=20^{\circ}$ would not significantly affect the nonlinear acoustic response of the material. Actually, a dramatic change in the material response occurs just upon passing the previous maximal angle of $\theta_2=20^{\circ}$: the intensity of the nonlinear signal variations, which is determined by the intensity of the weak-contact rearrangements, abruptly increases (figure~\ref{reactiv}). This pronounced effect of the packing orientation indicates the dominant role of gravity in the weak-contact rearrangements. The strong increase in the noise intensity after surpassing the previous maximal tilt angle is also remarkable (see figure~\ref{reactiv}). Similar behaviors are observed in other overlapped ranges of the tilt angle. As described in section~\ref{sec:surface}, the superficial rearrangements also demonstrate simultaneous reactivation which evidently contributes to the increased noise. 
%These observations are in good agreement with the results obtained on granular packings in a slowly rotating drum \cite{Kabla2005}. 
%The ageing of on-surface rearrangements  for tilted fresh-prepared packings in a limited angular range and their reactivation for larger angles, will be discussed in more detail elsewhere.

\subsection{Formation of a new weak-contact subnetwork and ``mirror-type'' internal avalanches of the weak-contact network near zero angle}
\begin{figure}[htbp]
\center
\includegraphics*[width=0.6\textwidth]{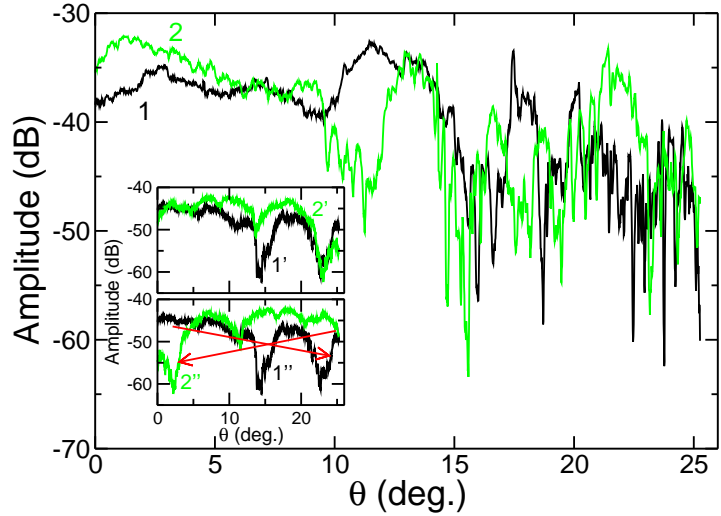}%{mirror_exmpls2.pdf}
\caption{Examples of the symmetrical  ``mirror-type'' evolution of the weak-contact network
(bead diameter $d = (3 \pm 0.1)\ \mbox{mm}$). %Curve 1 is for  the first tilt from zero to near-critical angle of a fresh prepared packing and curve 2 for the return. 
\textcolor{black}{The main panel shows superimposed curve 1 for the first
tilt from zero to near-critical angle of a freshly prepared packing and the very similar curve 2
for the return. The upper inset shows analogously superimposed curves 1' and 2' for the second
up-and-down cycle of tilting in another similar experiment. In the
lower inset, for the same pair of the curves, the latter curves have been replotted but the return curve (nammed 2'') is now inverted (i.e., the
return angle is counted from the maximal value down to zero). The resultant two-branch curve is rather similar to antisymmetric hysteretic loops observed in~\cite{Deboeuf2005}
for other physical characteristics related to the state of the contacts.}}%
\label{mirror-exp}%
\end{figure}

Despite similarities in the responses of the surface and of the bulk to consecutive tiltings, the surface activity and the weak-contact rearrangements differ in some respects. The surface activity decreases strongly during consecutive forth-and-back tiltings of the pile. The bulk rearrangements of the weak-contact network can behave differently, especially when the pile is tilted in the entire angular range, from zero to a near-crital angle. Figure~\ref{mirror-exp} shows as an example the superimposed records of the $F$ variations during titling from zero to a nearly critical angle of $\theta_m = 26^\circ$ and back from $26^\circ$ to zero. If the angle is counted from the initial position of the packing (i.e., from $\theta_m = 26^\circ$ for the return to $0^\circ$), the two curves exhibit a striking similarity. For the descending phase, a pronounced activity of the weak-contact rearrangements is observed for the angles close to the apparently stable horizontal position of the packing. This observation can be tentatively explained as follows. When coming closer and closer to  the nearly critical angle, the configuration of the weak-contact fraction gradually adapts itself to the changing orientation of gravity. In particular, such contacts with "mobilized" dry friction exhibit jumps into new, more stable positions as seen from 2D numerical simulations \cite{Staron2002,Staron2006}. For the reorganized weak-contact network, the new stable configuration  corresponds to angles close to the critical one. When the angle of the pile decreases back to to zero during the second half-cycle, gravity pushes such contacts out of their stable configuration formed near the critical angle. \textcolor{black}{Therefore, closer  to zero angle, such contacts become mobilized. They loose their stability  and exhibit a kind of "internal avalanche". They rearrange themselves back to more stable positions corresponding to near-horizontal orientation of the pile.  Thus the weak-contact network exhibits very significant modifications. Note that this is not the the case for the network of averagely loaded contacts, which mechanically supports the "visible" structure of the pile and ensures fairly stable conditions for the linear propagation of the sounding signal (i.e., the stable value of the elastic modulus)}. For a few-degree excursions, ageing of weak-contact rearrangements occurs. By contrast, for sufficiently large excursions of the tilt angles (especially between zero and near-critical angles), the weak-contact network exhibits unceasing rearrangements. We verified that these features are observed for grains of different types whose diameter varies from $1.8$ to $3.1\ \mbox{mm}$ (correspondingly the inertia effects differ by over $3.5$ times). Such "mirror-type" rearrangements in the bulk, which are seen to show quite good symmetry (figure~\ref{mirror-exp}), are fairly reproducible. They do not tend to cease even after $10-15$ tilt cycles in contrast to the surface activity.  However, during the very first tilts of freshly prepared packings, the variations of $F$ normally exhibit finer structure (as shown in the main panel of figure~\ref{mirror-exp}). For subsequent mirror-type rearrangements of the weak-contact network, larger scale structures are essentially seen (insets of figure~\ref{mirror-exp}).
 
 \begin{figure}[htbp]
\center
\includegraphics*[width=0.6\textwidth]{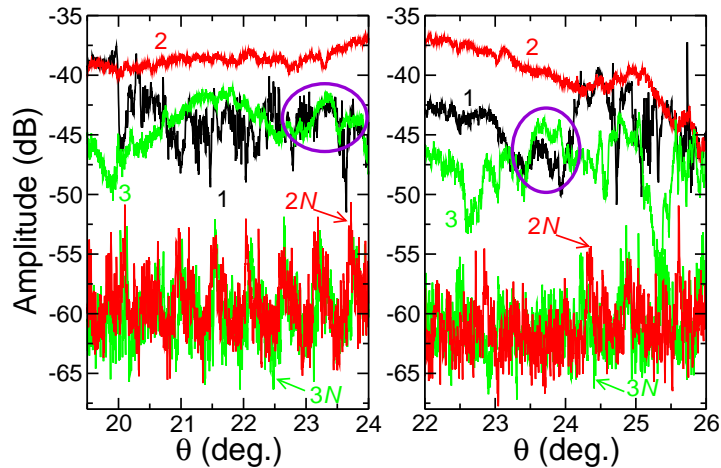}%{mem_short_new.pdf}
\caption{Examples of relaxation of the weak-contact network for two subsequent angular ranges of tilting (bead diameter $d = (2 \pm 0.1)\ \mbox{mm}$).  Curves $1$ are for the first upward tilt in the angular range. Aged curves $2$ are recorded after three (left panel) and two (right panel) consecutive up-and-down tilts. Curves $3$ are obtained after $20\mbox{ min}$ (left panel) and $45\mbox{ min}$ (right panel) of rest at the minimal angle before the next tilt.  The noise curves $2N$ and $3N$ correspond to curves $2$ and $3$, respectively. }%
\label{memory}%
\end{figure}

\textcolor{black}{It is worth noting that if the origin for the return angle is not shifted to zero but counted from the maximum back to zero value, the
ascending and descending branches form a kind of antisymmetric hysteretic loop. One
can mention a qualitative resemblance to antisymmetric hysteretic loops numerically
simulated in~\cite{Deboeuf2005} (e.g., figure 5 therein) for forth-and-back tilting of 2D granular packings.
Note that the quantity reported in~\cite{Deboeuf2005} (the density of critical contacts) is closely related to the experimentally observed
hysteretic behavior, since the variations of the nonlinear acoustic signal are determined by the instability (breaking) of the weak critical contacts.}

%Although the simulation relates to a significantly different characteristic of the contacts
%(density of critical ones), evidently it is closely related to the experimentally observed
%hysteretic behavior, since it is the instability (breaking) of the weak critical contacts,
%which determines the discussed variations of the nonlinear-acoustic signal.}

\subsection{Relaxational restoration of aged weak-contact network}
%Examples of "memory" of granular systems about the previous state and/or regimes of modification (e.g., %compaction regimes \cite{Rennes,Jaeger}) are known in literature. 
Deboeuf et al. \cite{Deboeuf2005,Deboeuf2003} observed the relaxation toward a static packing in terms of critical contacts by stopping the rotation. Their 2D numerical simulations show that the density of critical contacts is a dynamical response function to the actual loading. It vanishes after a transient dynamics. The critical contacts likely give rise to microplasticity. In previous experiments \cite{Zaitsev2008}, the nonlinear acoustic technique revealed active structural rearrangements even in the motionless granular material up to $10-30$ min after avalanche. It was thus tempting to search for such kinds of relaxation phenomena in our ageing experiment.
Despite the essential ageing of the weak-contact network towards an indifferent equilibrium state observed for moderate angular excursions (figures~\ref{ageing} and \ref{mirror-exp}), the weak-contact network 
%also keeps some memory about its initial configuration and even 
tends to spontaneously relax towards the pre-aged structure if the packing remains at rest for sufficiently long time. This statement is illustrated by figure~\ref{memory}. The left part corresponds to a range of tilt angles of $18^\circ-24^\circ$, and the right part to the subsequent range of $20^\circ-26^\circ$. Curves marked $1$ are obtained for the first upward tilt in the angular range indicated. They exhibit pronounced variations due to weak-contact rearrangements (in particular, curve $1$ in the left plot in figure~\ref{memory} is the same as curve 1 in figure~\ref{reactiv}) 

 Significantly smoother curves $2$ in both parts of figure~\ref{memory} correspond to aged structures of the weak-contact network obtained after three and two up-and-down cycles for the left and right panels, respectively. Before recording curves marked $3$, the packings remained $20-45\mbox{ min}$ at complete rest near their respective minimal angles. Instead of showing changes similar to (or even smaller than) those found for previous cycles, the relaxed curves $3$ reveal significantly reactivated rearrangements. Their variations are similar to those of curves $1$.
In both cases, the weak-contact configurations restore themselves during the rest periods, into states which are close to their states before the first up-tilting. The restored configurations are however not such that curves $3$  reproduce entirely the shapes of curves $1$. We observe nevertheless similarities between some parts of curves $1$ and $3$ (e.g. the encircled ones), as well as in the angular positions where abrupt variations occur (intensity jumps). Some features of the aged network appear thus to be restored.
 
The possibility of such restoration can be understood since the weak-contact rearrangements do not imply macroscopic displacements of the grains, such that almost all contacts remain in positions very close to their initial ones until the onset of the macroscopic avalanche occurs. This ensures the possibility for  parts of the weak contacts to return from the aged metastable configuration to the equilibrium positions corresponding to smaller tilt angles. It is important to emphasize that the acoustic signal was switched off during the rest periods. Therefore, the restoration of the aged configuration is attributed to the intrinsic dynamics of the packing under the influence of gravity and thermal fluctuations. The latter factors evidently suffice to overcome energy barriers resulting from the nanometer scale distances between the ruptured and slightly separated interfaces at the weak contacts. Figure~\ref{memory} shows that this configuration can be spontaneously restored to a considerable extent, even after significant disturbance, if the packing is returned to the initial angle and if it is maintained at rest for a sufficiently long time.  
 
It is also interesting to compare the levels of its own acoustic noise in the tilted packing for the aged and then restored configurations. The own material noise within a $128 \ \mbox{Hz}$ bandwidth around $2816 \ \mbox{Hz}$ (as in figure~\ref{reactiv}) is shown in figure~\ref{memory} by curves marked $2N$ and $3N$ which correspond to curves $2$ and $3$. The noise becomes $3-6\ \mbox{dB}$ more intense after the rest period. Comparison with the much stronger noise in freshly prepared packings (figure~\ref{reactiv}) indicates that the noise is initially dominated by the superficial activity.

\section{Conclusions}
Optical measurements and non-linear acoustic technique were combined to study destabilization in tilted granular packings. The resulting grain rearrangements at the surface of the pile as well as the  modifications of the weak-contact network can be followed in that way. 
These two signals %on-surface rearrangements of beads and modifications of the weak-contact network 
exhibit weak, although discernible correlations at least for freshly prepared granular piles.  
%The performed complementary visual and nonlinear acoustic observations confirmed that on-surface rearrangements of beads for tilted granular piles and modifications of the weak-contact network (monitored using the nonlinear acoustic approach) can exhibit weak, although 
%significant 
%discernible correlations, but only for fresh prepared granular piles. 
These correlations seem to disappear when the piles are inclined repeatedly forth and back. Additional more detailed statistical studies of the on-surface displacements, of the weak-contact network modifications and of the own acoustic noise in granular piles are planned.
The reported experiments constitute evidence of the existence of ageing and memory effects in 3D packings slowly driven to their maximum angle of stability. During inclination cycles, the activity of a granular packing decreases (after 3-5 cycles) and reaches a stationary state. 
%Both measurements evidence the existence of ageing and memory effects in 3D packings slowly driven to their maximum angle of stability. Nevertheless the signature of ageing is somewhat different.
   %Experiments using non-linear acoustical techniques evidence similar effects in the bulk of the packing \cite{vlad2}. They will allow us to compare the respective roles played by the weak and strong contact networks in the 
   %relaxation process 
  %dynamics and consequently their roles in the unjamming transition. 
%  The reported experiments constitute evidence of the existence of ageing  in 3D packings slowly driven to their maximum angle of stability.
%During inclination cycles, the surface activity of a granular packing decreases (after 3 cycles) and reaches a stationary state.  The treatment of the modification of the weak-contact network is more subtle. 
%Similarly to what is observed for the surface activity, for moderate (within $6^\circ-8^\circ$) excursions of the forth-and-back tilting, the configuration of the weak contacts demonstrates pronounced ageing towards a new quasi-indifferent equilibrium state.\\
%
For cycles performed in overlapped ranges of the tilt angles, the aged configuration presents reactivation of the rearrangement activity upon surpassing the previous maximal tilt angle. This is observed for the weak-contact network modification as well as for surface activity.
%This fact is intuitively expected and resembles the similar situation for the on-surface displacements of the grains.\\
 For larger excursions, especially for inclinations in the entire range from zero to near-critical angles, very peculiar rearrangements occur in the weak-contact fraction instead of ageing towards an indifferent configuration. These rearrangements appear as ``mirror-type internal avalanches'' when the packing is tilted up to an almost critical angle and then returned to the horizontal position. For the descending half-cycles, the configuration of weak contacts rearranged during upward tilting  loses its stability closer to zero angle producing a kind of ``mirror" symmetry of the weak-contact avalanches. This effect does not exhibit any trend toward ceasing and is very well reproducible for multiple forth-and-back cycles, unlike irreversibly aged on-surface displacements.
Finally, it is found that even apparently aged configurations of the weak-contact network can exhibit a peculiar slow restoration towards the configuration it had before ageing if the packing remains at rest during tens of minutes. These results have a deep relation to earlier reported relaxation effects of other types found for granular materials \cite{Deboeuf2005,Bocquet1998}. In a broad sense, the  experimental results presented support and complement the conclusion following from numerical simulations in 2D \cite{Radjai1998, Deboeuf2005,Staron2002, Staron2006} and from earlier acoustical \cite{Zaitsev1995,Tournat2004} results that granular materials can be viewed as a composition of essentially independent fractions of strong and weak contacts with radically different properties.\\

Our results suggest that the non-stationary dynamics of an inclined grain packing shares relaxation properties and memory effects with granular media subjected to different kinds of mechanical perturbations (for instance, tapping \cite{Josserand2000,Ribiere2007,Ribiere2005b}).  
%This process displays some similarities with granular compaction under gentle vertical tapping where both long time relaxation~\cite{Ribiere2007} and memory effects~\cite{Josserand2000} are observed.  
%This process may be linked with the process of compaction observed in granular matter under vertical tapping \cite{Ribiere2007} where relaxations are observed to occur on long time scales . 
ageing effects stabilize the packing. They emphasize the major role played by the orientation of gravity in the reorganization of contact networks. In particular, the activity of the packing was shown to depend strongly on the past of the system. This ageing effect has to be considered in parallel with the influence of the initial preparation of the packing (for example the initial packing fraction). Thus, information on that past is in some way stored in a specific texture of contact between grains. 
\section{Acknowledgments} The study was supported in parts by ANR grants STABINGRAM, ANR-05-BLAN-0273, NT-05-3-41489, and RFBR-PICS grant No 09-02-91071-CNRS. V.Z. acknowledges the Universities of Maine and Rennes~1 for obtaining invited-professor grants.
We thank A. Faisant and E. Brasseur for technical help, D. Bideau, Ph. Boltenhagen and D. Iudin for helpful discussions.\\

\bibliographystyle{unsrt}
\bibliography{biblio_prec}

\end{document}